\documentclass[alpha-refs,english]{RBCA_v2.0}

\usepackage{longtable}
\usepackage{ragged2e}

\title{Virtual Reality Applications in Software Engineering Education: A Systematic Review}

\titleother{Título do artigo em inglês}

\author[1]{Gustavo Vargas de Andrade} 
\author[1]{André Luiz Cordeiro Gomes}
\author[1]{Felipe Rohr Hoinoski}
\author[1]{Marília Guterres Ferreira}
\author[1]{Pablo Schoeffel}
\author[1]{Adilson Vahldick}

\affil[1]{Universidade do Estado de Santa Catarina, GameLab}

\authnote{\authfn{1}gupontoandrade@gmail.com; andre.cordeiro0612@gmail.com; felipe.rh@edu.udesc.br; marilia.gf@udesc.br; pablo.schoeffel@udesc.br; adilson.vahldick@udesc.br}

\runningauthor{First et al.}    

\papercat{Original Paper}

\jvolume{11}           
\jissue{3}             
\jyear{2019}           
\jmonth{November}      

\firstpage{1}
\jid{9999}             

\jrec{yyyy-mm-dd}      
\jrev{yyyy-mm-dd}      
\jacc{yyyy-mm-dd}      

\begin{document}

\begin{frontmatter}
	
\maketitle
\thispagestyle{empty}

\begin{Abstract} 
Requirement Engineering (RE) is a Software Engineering (SE) process of defining, documenting, and maintaining the requirements from a problem. It is one of the most complex processes of SE because it addresses the relation between customer and developer. RE learning may be abstract and complex for most students because many of them cannot visualize the subject directly applied. Through the advancement of technology, Virtual Reality (VR) hardware is becoming increasingly more accessible, and it is not rare to use it in education. Little research and systematic studies explain the integration between SE and VR, and even less between RE and VR. Hence, this systematic review proposes to select and present studies that relate the use of VR applications to teach SE and RE concepts. We selected nine studies to include in this review. Despite the lack of articles addressing the topic, the results from this study showed that the use of VR technologies for learning SE is still very seminal. The projects based essentially on visualization. There are lack of tasks to build modeling artifacts, and also interaction with stakeholders and other software engineers. Learning tasks and the monitoring of students’ progress by teachers also need to be considered.
\end{Abstract}

\begin{keywords}
Educational software; Requirements engineering; Software engineering; Virtual reality; 
\end{keywords}

\begin{resumo} 
Engenharia de Requisitos (ER) é um processo de Engenharia de Software (ES) que define, documenta e mantém os requisitos de um problema. É um dos processos mais complexos da ES porque aborda a relação entre cliente e desenvolvedor. O aprendizado de ER pode ser abstrato e complexo para a maioria dos estudantes porque muitos deles não conseguem visualizar o assunto aplicado. Através do avanço da tecnologia, o hardware de Realidade Virtual (RV) está se tornando cada vez mais acessível, e não é raro utilizá-lo na educação. Poucas pesquisas e estudos sistemáticos explicam a integração entre ES e RV, e ainda menos entre ER e RV. Esta revisão sistemática propõe selecionar e apresentar estudos que relacionam o uso de aplicações de RV para ensinar os conceitos de ES e ER. Selecionamos nove estudos para incluir nesta revisão. Apesar da falta de artigos abordando o tema, os resultados deste estudo mostraram que o uso das tecnologias de RV para o aprendizado de ES ainda é muito seminal. Os projetos se baseiam essencialmente na visualização. Faltam tarefas para construir artefatos de modelagem, e interação com as partes interessadas e outros engenheiros de software. As tarefas de aprendizagem e o monitoramento do progresso dos alunos pelos professores também precisam ser considerados.
\end{resumo}

\begin{palavras_chave} 
Engenharia de requisitos; Engenharia de software; Realidade virtual; Software educacional;
\end{palavras_chave}

\end{frontmatter}

\section{Introduction}
Using Virtual Reality (VR) in education can be considered a natural evolution in using technology to support learning \citep{pantelidis}. VR allows novel forms and methods of visualization and interaction, providing the student with more proximity and depth in the observation and examination of objects and processes. Students feel motivated and challenged to walk and interact in a 3D environment, and even more with the possibility of changing that environment \citep{pantelidis}. Because of these qualities, the idea of combining VR with Software Engineering (SE) to make abstract subjects tangible arises. In the past, one disadvantage of VR was related to the cost. This problem has already been overcome because, with the latest generation of smartphones and a foldable cardboard display, users can reach these immersive environments \citep{tori}. However, according to \cite{hillmann}, the quality of VR hardware and software is not yet good enough for a mass market.

Requirements elicitation and specification are the first phase of software development. Problems at this stage account for over 50\% of software errors \citep{avila}. Therefore, the quality in elicitation and specification of requirements is critical to the success of a software project. For this reason, there is a need to prepare students for Software Engineering courses, offering them the opportunity to experience this process during their training. Universities may not be creating enough skills for future engineers to perform this task \citep{thiry}. It is still necessary for the academic environment to develop maturity and practices with real projects for students to develop the skills to conduct Requirement Engineering (RE) \citep{romero}. However, this task depends on external factors, in which real or fictitious customers are available to assist students.

In this article, we carry out a systematic review of the existing VR applications to SE education and highlight the immersive aspect of VR to answer seven research questions aiming to identify the features and resources developed in these studies. Although our initial focus was on studies in RE, we have expanded the study to find features from other applications related to SE that can contribute to the teaching of RE. Our goal is to contribute to the existing body of knowledge on the application of VR for educational purposes in RE. We did not find reviews involving the use of VR in the SE context. Many reviews that we found claimed to have used VR, but in reality, they used just 3D environments (such as a 3D rendered on a computer or smartphone screen). It was very difficult to base and answer the proposed questions because we found few studies that include both VR and SE education concepts. We can consider this paper a pioneer in addressing the theme.

\section{Background and related work}
Following the definition from \cite{biocca}, VR results from the sum of hardware and software seeking to give users the immersive feeling that they are in another reality. \cite{wang} define VR as the visualization technique referred to a pure virtual presence. They also split VR into five categories: Desktop-Based VR, Immersive VR, 3D Game based VR, BIM-Enabled VR, and Augmented Reality. The present paper focuses on reviewing the use of immersive VR, which is the category of visualization that requires special hardware, such as head-mounted displays (HMDs) and Projection-based displays (PBDs) \citep{feng}. HMDs comprise goggles attached to a helmet that displays images for each eye and PBDs are related to creating immersive rooms by displaying images in one or more walls \citep{lanzagorto}.

To give the user high immersion, a newer generation of HMDs is being used \citep{radianti}. In the past, the use of HMDs was expensive and uncomfortable \citep{kerawalla}. Users can overcome the problems that have arisen with price when they have access to new generations of smartphones, allowing a greater possibility of accessing VR environments \citep{tori}. One of the key problems arising from the physical problems is cybersickness, which can mainly affect people who are not used to VR games \citep{jensen}.

It is common to have studies applying VR in education \citep{abdullah, gutierrez, vretos}. Usually, its conclusions claim the results are positive, somehow fostering and increasing student’s productivity. The literature review presented by \cite{radianti} searched how VR applications are applying for high educational purposes. To do that, the authors analyzed and categorized 38 papers into eight questions. We used many definitions and categorizations defined by this review to base our questions, such as immersive VR designs and validations designs. That review concludes that the variety of papers using VR shows that there is a high expectation in the content, but the maturity of the studies still can be questionable because few of them showed how their experiments can be implemented in the teaching curriculum.

Software are computer programs and associated documentation, which can be developed for a particular customer, or for the general market \citep{sommerville}. SE is the engineering discipline responsible for software production, from initial conception to operation and maintenance. SE is fundamental to society because it affects almost every aspect of our lives, changing the way we, as humanity, deal with commerce, culture, and our everyday activities \citep{pressman}.

\begin{figure}[bt!] 
\centering
\includegraphics[width=\linewidth]{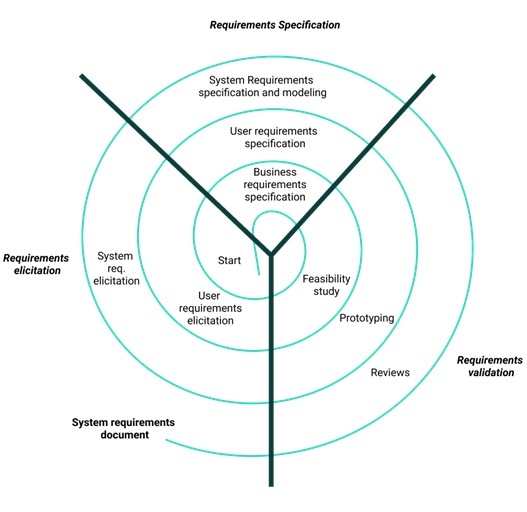}
\caption{Requirements engineering process. \citep{sommerville}}\label{fig:req_eng_process}
\end{figure}

\cite{sommerville} defines requirements as descriptions of the services that a system should provide and the constraints in its operation. RE can be considered the process of finding out, analyzing, documenting, and checking these services and constraints. That means RE is the first step in software development. It handles the communication between customers and the development team. It might seem like a simple process, but usually, customers and users are not familiar with SE processes, which may cause problems of scope, understanding, and volatility \citep{pressman}. \cref{fig:req_eng_process} describes how the RE process works. \cite{sommerville} separated RE into three key activities: requirements elicitation, specification, and validation. In requirements elicitation, the engineer must discover the requirements by interacting with the stakeholder. In requirements specification, the engineer must convert the collected requirements into a standard form, such as a requirements specification document. In requirements validation, the engineer must check if the requirements define the system that the user asked for.

Designing an approach to convey RE concepts, which must encompass practical experience, is considered a hard task for technology courses \citep{portugal}. \cite{sikkel} presented some problems for those who are not teachers, such as the difficulty in validating the requirements in a practical way. The study also claims that in university, students are used to a perfect, self-contained problem description. There are studies describing the integration of students with industry, but usually students and instructors assume different roles to enact the practical experience \citep{portugal}.

\cite{santos} presented a review where they merged evidence regarding the use of the RE process for VR systems and the VR contributions to the RE process. The results claimed few researches were describing the use of RE for VR, or VR to support the RE process. The review identified the use of many SE models and adapted them to the RE process for VR systems.

\section{Methodology}
\begin{figure*}
\centering
\includegraphics[width=.6\linewidth]{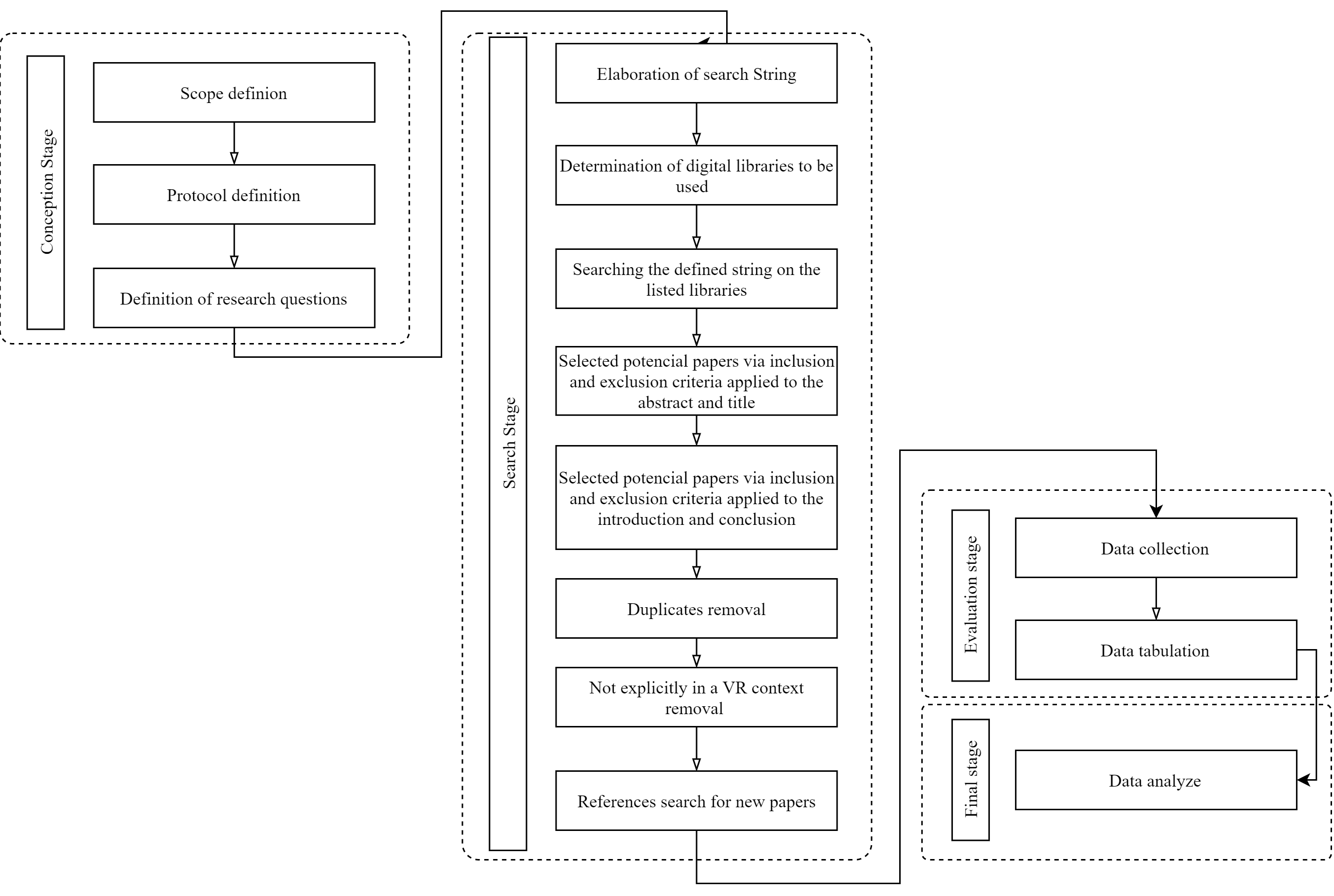}
\caption{Systematic Literature Review process. Adapted from
\cite{kitchenham}}
\label{fig:srl_process}
\end{figure*}

 Systematic Literature Review (RLS) is defined by \cite{kitchenham} as “[...] identifying, evaluating and interpreting all available research relevant to a particular research question, or topic area, or phenomenon of interest.” We conducted the overall review process of this research in four main stages as illustrated in \cref{fig:srl_process}. Therefore, the following sections will report the complete process.
 
 \subsection{Conception Stage}
 
 The first step to conduct the SLR was the conception stage, by defining the scope, protocol, and research questions. The protocol development is a vital part of any SRL because it allows other researchers to recreate the same review and “reduce the possibility of researcher bias” \citep{kitchenham}. Being aware of this study, which is to review the use of VR in the educational application of SE and RE, we defined the research questions as described in \autoref{tab:tab1}.
 
\begin{table}
  \caption{Research questions of the study}
  \label{tab:tab1}
  \setlength{\tabcolsep}{10pt} 
\renewcommand{\arraystretch}{1.5}
 \begin{tabular*}{\linewidth}{p{3.5mm} p{7.4cm}}
 
    \toprule
   \# & Research Question (RQ)\\
    \midrule
   
RQ1 & What subjects of SE or RE were simulated in VR
environments?\\
RQ2 & What were the VR features in SE or RE teaching
environments?\\
RQ3 & How were the students' interactions performed?\\
RQ4 & What and how were the developer-customer-user
interactions?\\
RQ5 & Which Unified Model Language (UML) diagrams
were adopted and how they were used and designed
by students in VR systems?\\
RQ6 & Which were the target hardware of the studies?\\
RQ7 &How were the virtual teaching environments in SE
or RE evaluated?\\
  \bottomrule
\end{tabular*}
\end{table}

 We will better explain the questions and describe how each found paper answers the questions, in the result section.
 
\begin{figure}[bt!] 
\centering
\includegraphics[width=\linewidth]{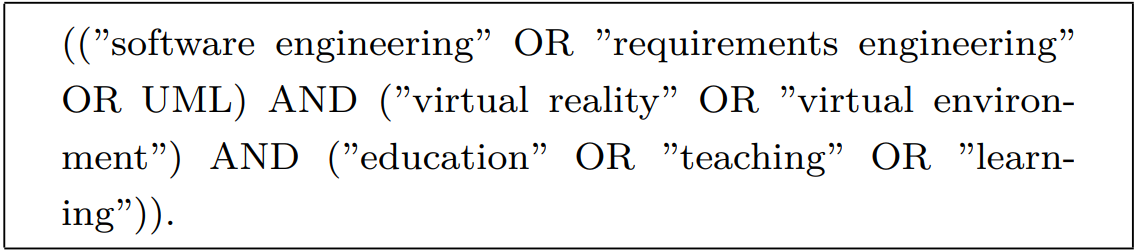}
\caption{Search string defined for the scope of this work}\label{fig:search_string}
\end{figure}

 To find the relevant studies, we defined the search string (search stage shown in \cref{fig:srl_process}) following the question structure (population, intervention, outcomes, and experimental designs) proposed by \cite{kitchenham}. We illustrate the keywords used on the string in \cref{fig:search_string}. We performed the investigation manually using research databases strengthened for its high range and quality of publications \autoref{tab:tab2}).
 
\begin{table}[ht]
  \caption{Knowledge bases used for data research}
  \label{tab:tab2}
   \setlength{\tabcolsep}{10pt} 
\renewcommand{\arraystretch}{1.5}
 \begin{tabular*}{\linewidth}{p{2cm} p{5.9cm}}
    \toprule
  Acronym & Name\\
    \midrule
   
ACM & Library Association for Computing Machinery\\
IEEEXPLORE & IEEE Xplore digital library\\
ResearchGate & ResearchGate\\
Scielo & Scientific Electronic Library Online\\
ScienceDirect & ScienceDirect\\
Scopus & Scopus\\
  \bottomrule
\end{tabular*}
\end{table}

The search string returned a different quantity of studies on each site. Then, we defined inclusion and exclusion criteria (\autoref{tab:tab3}) to filter the primary studies selected, aiming to choose those that were part of this study's goal.

\begin{table}[ht]
  \caption{Inclusion and exclusion criteria}
  \label{tab:tab3}
   \setlength{\tabcolsep}{10pt} 
\renewcommand{\arraystretch}{1.5}
 \begin{tabular*}{\linewidth}{p{1.3cm} p{6.6cm}}
    \toprule
Type & Criteria\\
    \midrule
   
Inclusion & Environments for teaching or simulating SE
and RE in VR\\
Exclusion & SE processes for VR development \newline
Studies in the field of medicine and robotics\newline
Virtual environments unrelated to SE
context\\
  \bottomrule
\end{tabular*}
\end{table}
 
 After defined the inclusion and exclusion criteria, as well as the search terms, we carried out the systematic review following the previously planned steps.
 
 \subsection{Search Stage}
 
\begin{figure}[h]
  \centering
  \includegraphics[width=5cm,height=14cm]{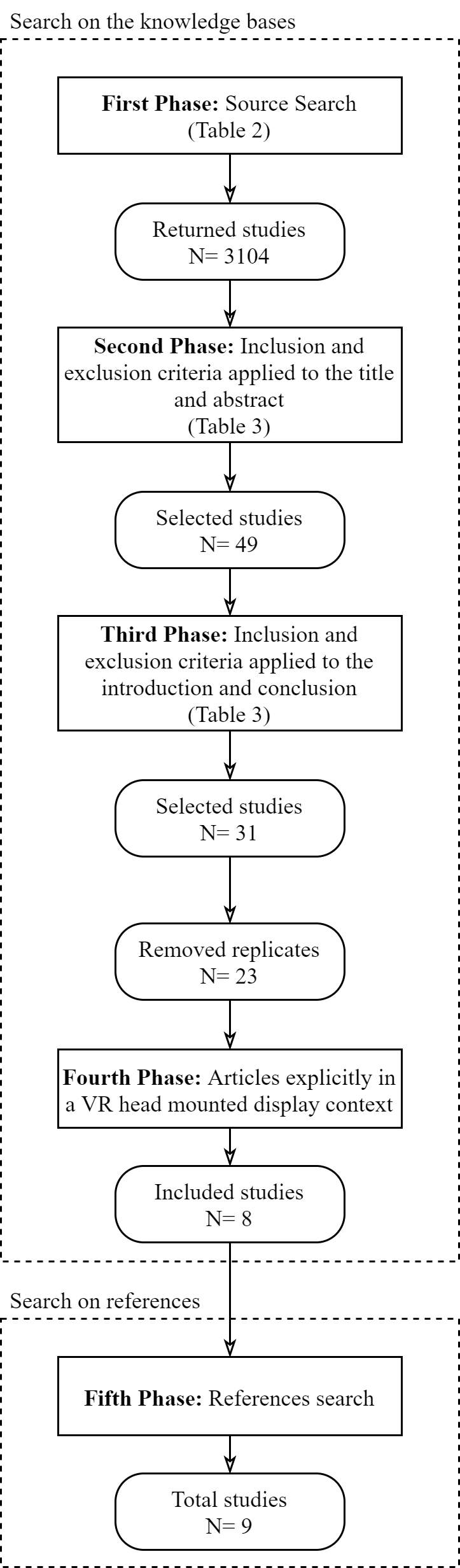}
  \caption{ Research phases and number of included studies}
  \label{fig:res_phases}
\end{figure}

This stage of the research process aimed to build a list of articles that will be the object of study in the next stage. \cref{fig:res_phases} shows the overall view and the number of resulting studies in each phase of the research process, and the same total number of studies are separated by knowledge bases in \autoref{tab:tab4}.

First, the search string (\cref{fig:search_string}) was applied on each research base listed in \autoref{tab:tab2}, returned the total amount of related works as a preliminary result (First Phase: 3104 studies - \autoref{tab:tab4}). Then, we read each paper title and excluded the ones that do not explicitly address the subject, according to the inclusion and exclusion criteria (\autoref{tab:tab3}), resulted in 49 studies in the second phase. Thereafter, we read the Abstract, Introduction, and Conclusion sections, resulted in the removal of eighteen papers. We removed the replicated papers (Third Phase: 23 studies). Finally, we read each selected paper in its totality to identify a misunderstanding regarding the concept of VR environments. Some papers described VR environments without the use of dedicated hardware, i.e. the user just had to be inserted into a 3D environment through the computer screen, instead of defined VR by \cite{biocca} as the sum of hardware and software. Thus, we selected only the studies that explicitly used a VR head-mounted display, resulting in a list of eight studies in the fourth phase.

As an additional step, we looked for papers referenced by the eight selected studies, considering the inclusion and exclusion criteria, to avoid that some relevant papers were not included because they are outside the research bases. In this step, we added one new study resulting in the total of nine studies on VR learning environments applied in software and requirements engineering (fifth phase).

\begin{table}[ht]
  \caption{Number of studies by research database}
  \label{tab:tab4}
   \setlength{\tabcolsep}{10pt} 
\renewcommand{\arraystretch}{1.5}
 \begin{tabular*}{\linewidth}{p{4cm} p{3cm}}
    \toprule
Source & Number of studies\\
\midrule
ACM &1097\\
IEEEXPLORE &182\\
Researchgate& 100\\
Scielo &1\\
Science direct& 1281\\
Scopus &443\\
\midrule
Total&3104
\\
  \bottomrule
\end{tabular*}
\end{table}
Each phase was performed by one author and repeated by two other researchers to ensure consistency in the outcomes. If some work was not included or excluded, all the researchers discussed it until they reached a consensus.

\subsection{Evaluation Stage}

Once we finished the search stage, we collected and tabulated the relevant information of each nine studies found, to answer each of the research questions listed in \autoref{tab:tab1}. It will organize the details about the results for each paper in the next section, composing the Last Stage of this RSL (shown in \cref{fig:srl_process}). \autoref{tab:tab5} lists the nine studies.

\begin{table*}[ht]
  \caption{List of selected studies}
  \label{tab:tab5}
   \setlength{\tabcolsep}{10pt} 
\renewcommand{\arraystretch}{1.5}
 \begin{tabular*}{\linewidth}{p{4.8cm}p{12.2cm}}
    \toprule
Study & Features\\
\midrule

\RaggedRight{\cite{akbulut}} & Describes a system (VRENITE) for learning sorting algorithms in which students experience the elements to be sorted.
\\ 
\RaggedRight{\cite{elliott} }& The paper describes the two applications, the “RiftSketch” and “Immersion”. The first one presents an application that lets the user program in a VR world, writing the codes and seeing the real-time feedback of them. The other  aims to give an alternative for code reviewers, bringing codes in tables in a VR environment. 
\\ 
\RaggedRight{\cite{fittkau}} & Describes the benefits for understanding the software system from different perspectives, including VR, using the software (ExplorViz) in the research, mapping the systems architecture, and using HMD specific technologies to design the software architecture in VR environments.\\
\RaggedRight{\cite{merino}}&As \cite{oberhauser} and \cite{vincur}, this paper describes an application that allows the user to see program code structures like buildings.
\\
\RaggedRight{\cite{oberhauser}}& As \cite{merino} and \cite{vincur}, this paper abstracts software code structures as buildings in a virtual world.
\\
\RaggedRight{\cite{vincur}}&As \cite{merino} and \cite{oberhauser}, this paper describes an application that presents software code structures as buildings in a VR environment.
\\
\RaggedRight{\cite{ochoa}}&Describes a teaching application that approaches software requirements through virtual reality UML modeling.
\\
\RaggedRight{\cite{tanielu}}&Describes the use of VR to support the teaching of Object-Oriented Programming (OOP). To do that, the researchers took OOP concepts and abstracted them to constructions, which resulted in an application called OOPVR.\\
\RaggedRight{\cite{zhao}}&Describes a VR app to help students understand the manufacturing process. It comprises presenting to the user seven stations, in which he will need to assemble a manufactured product. At the first station, the user can analyze the requirements of the request, and on the second station, he can select which pieces he will need to craft the product. The five subsequent stations were for the assembly of the vehicle.\\
  \bottomrule
\end{tabular*}
\end{table*}

\section{Results}
\begin{figure}[h]
  \centering
  \includegraphics[width=\linewidth]{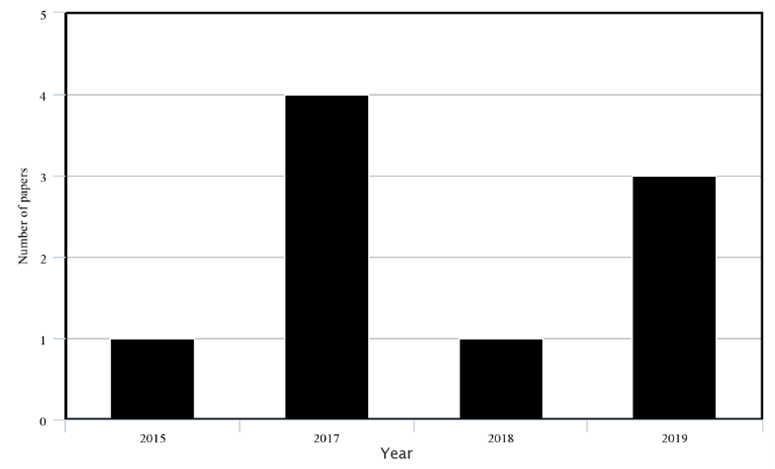}
  \caption{ Relation of the number of papers published for each year}
  \label{fig:papers_year}
\end{figure}

The selected studies were from 2015 to 2019. This emphasizes the idea that studies in the SE area that work with virtual reality are quite recent, as shown in \cref{fig:papers_year}.

\subsection{RQ1: What subjects of SE or RE were simulated in VR environments?}

The area of software engineering is subdivided into several others, and some of the selected papers were related to other SE content requirements engineering. \cite{fittkau, merino, oberhauser, tanielu} and \cite{vincur} present environments of programming course, and are related to the software visualization, in which the user is a floating spectator who can observe the classes and packages of her/his system in a more playful environment, using the abstraction of cities and buildings to understand the connections of an application. \cite{fittkau, merino, oberhauser} and \cite{vincur} are also applications focused on software architecture.
\cite{elliott} present a VR programming system, in which the student has real-time feedback on what is programmed, being able to instantiate trees, birds, and leaves that appear in a virtual world. \cite{akbulut} show an application that aims to teach software concepts controlled by a main user (possibly a teacher) to other users with a VR display. The students cannot interact with the software.
\cite{ochoa} and \cite{zhao} deal with requirements engineering. In the application presented in \cite{ochoa}, the user connects floating UML diagram elements, which are usually performed in two-dimensional environments. \cite{zhao} aims to train people in the stages of the manufactured production process, using the analysis of functional requirements for the creation of a product requested by a customer.
According to the subject covered by each paper, we define three categories: Programming, Requirement, and Software Architecture, which are shown in \autoref{tab:tab6}. Programming refers to any VR application related to code and to construct algorithms. Requirement is referred to as any VR application related to RE. Software architecture is related to any VR application related to structural decisions and management.

\begin{table} [ht]

  \caption{Subject covered per study}
  \label{tab:tab6}
   \setlength{\tabcolsep}{10pt} 
\renewcommand{\arraystretch}{1.5}
 \begin{tabular*}{\linewidth}{p{3.5cm} p{5cm}}
    \toprule
Subjects & Studies\\
\midrule

Programming &\cite{akbulut} \newline\cite{elliott} \newline\cite{tanielu}\\
Software Architecture &\cite{fittkau} \newline\cite{merino} \newline\cite{oberhauser} \newline\cite{vincur}\\
Requirement &\cite{ochoa} \newline\cite{zhao} 
\\
  \bottomrule
\end{tabular*}
\end{table}

\subsection{RQ2: What were the VR features in the SE or RE teaching environments? }

With this question, we want to identify how VR resources are used in teaching software or requirements engineering. The features are described and listed in \autoref{tab:tab7}.

\begin{table*} [!ht]

  \caption{VR features applied in SE or RE teaching environments.}
  \label{tab:tab7}
\setlength{\tabcolsep}{10pt} 
\renewcommand{\arraystretch}{1.5}
\begin{tabular*}{\linewidth}{p{1.3cm} p{15.3cm}}

 \toprule
Study & Features\\
    \midrule
   
\RaggedRight{\cite{akbulut}} &The teacher sets up the experiments by defining the size of array and the algorithm: selection sort, bubble sort, insertions sort, or merge sort. Each student, equipped with an iOS smartphone, plays the role of an array element to be organized. The rearrangement of the array is captured when students change their
position.\\

\RaggedRight{\cite{elliott}} &Riftsketch allows the user to code in JavaScript and see real-time scene updates, such as the creation of animals and plants. Immersion allows the user to review code inside the VR environment by representing methods as code fragments in squared blocks. Both applications use an HMD, Oculus Rift development kit 2, and a Leap Motion Controller for gesture recognition. \\

\RaggedRight{\cite{fittkau}}& The system allows users to view the system architecture (class, packages, and their connections) by moving, rotating, and zooming in on the virtual scope. Students visualize the architecture as UML diagrams. It used colored boxes to identify classes or packages. There are four types of visualization.\\

\RaggedRight{\cite{merino}}&CityVR is an application which uses VR to visualize object-oriented programming concepts using an analogy of buildings, such as \cite{fittkau}, \cite{ochoa}, and \cite{zhao}. The system represents each class as a building in the city and the districts as software packages and allows the user to inspect the source code of the class, pointing the controller at it, and pressing a button to select it. \\

\RaggedRight{\cite{oberhauser}}& The system allows the user to control altitude (up, down) using one HTC touchpad; control directions (left, right, forward, backward) using another HTC touchpad; select an object and change its color; visualize his Java application in the city metaphor, where the higher the building, the bigger it is within its package. The system shows a laser pointer for selection when the controller enters the view field and has a virtual keyboard that supports text input for searching, filtering, and tagging. When a controller is near to and pointing at the oracle (virtual tablet), the system changes the virtual controller mesh to a finger for intuitive interaction. \\

\RaggedRight{\cite{vincur}}&VRCity is an application which uses VR to visualize object-oriented programming concepts using an analogy of buildings. The system represents classes as buildings and methods as floors. It allows the user (1) to scale the buildings (zoom in and zoom out); (2) to browse source code directly in a virtual reality environment; (3) to change the color scheme to filter between static aspects (metrics and coupling), dynamic aspects (trace visualization using disharmony maps), evolution (contribution of authors and the change of static aspects in a specified period); (4) to select objects that could be grabbed or triggered. To grab an object, they simply push grip buttons, which is like the motion required to grab a real-world object. To trigger an action attached to an object, a user must push the trigger button. Scrolling is provided by the track pad button. To scroll up or down, a user just simply swipes his finger across the track pad area on the HTC Vive controller. \\

\RaggedRight{\cite{ochoa}}&The system allows the user of the HTC Vive Controller to: draw lines between models, move models, filter the element; spawn UML elements in hand, in other words, students could create software analysis models which are typically created (as use case models) and manipulate their elements. The system presents the elements in a digital menu with all available models and their elements. The requirements found throughout the text are at the user level.\\

\RaggedRight{\cite{tanielu}}&OopVR is an application which uses VR to visualize object-oriented programming concepts using an analogy of buildings. The system allows the user to: visualize classes as a blueprint of a house, view instances of a class in the form of houses, visualize a method in the form of rooms inside the house; view variables and their values in the form of boxes inside the room; visualize the entry point of a method as the shape of a bedroom door; visualize method arguments placed on windowsills outside a room; visualize the accessibility of the fields in the center of the house; visualize the allocation in memory like the land of a house; visualize the address in memory in the form of a house mailbox; visualize the encapsulations by solidifying the walls of the house.  \\

\RaggedRight{\cite{zhao}}&The paper describes an application where the student can see the visualization of the functional requirements of a customer's request, and the parts that the user can select for the construction of the selected object; interaction with objects; construction of objects; the user to teleport between seven stations related to the object's assembly process. To display that, the student uses HTC Vive, wireless controllers, and base stations for motion tracking.\\
  \bottomrule
\end{tabular*}

\end{table*}

By collecting these features, we will analyze and see some similarities in them, such as the absence to contemplate the entire process of creating and maintaining a software, but focusing on smaller, specific parts of the complete process. It also shows that interactivity with the user is a key principle in VR. In the next question, according to the interaction with the system, we will separate each VR application into one or more categories.

None of the studies described the definition of an instructional unit, nor evaluations or monitoring of students' progress.

\subsection {RQ3: How were the students' interactions performed?}
The review of \cite{radianti} conceptualized fourteen categories of educational design elements: realistic surroundings, passive observation, moving around, basic interaction with objects, assembling objects, interaction with other users, role management, screen sharing, user-generated content, instructions, immediate feedback, knowledge test, virtual rewards, and making meaningful choices. To answer our question, we retrieved the attributes of each application in research question 1 and classified them in at least the category quoted before, as shown in \autoref{tab:tab8}.

\begin{table*} [ht]
  \caption{Studies per interaction categories.}
  \label{tab:tab8}
   \setlength{\tabcolsep}{10pt} 
\renewcommand{\arraystretch}{1.5}
 \begin{tabular*}{\linewidth}{p{2cm} p{10cm} p{4cm}}
    \toprule
Category & Description & Related Studies \\
    \midrule

Moving around &Students can explore the virtual environment, whether by teleporting or flying.&  \cite{elliott}
\cite{merino}
\cite{oberhauser}
\cite{tanielu}
\cite{vincur}
\\
User-generated content &Students can create new content, such as 3D models, and upload that new content to the virtual environment. This design element also applies when user-generated content can be shared with other users so that they can use it in their virtual environment. This design element does not apply when students can only access virtual objects that were created by developers and provided by a library in the virtual environment.&  \cite{elliott} \newline \cite{merino} \newline\cite{oberhauser} \newline\cite{tanielu} \newline\cite{vincur}
\\
Assembling &Students can select virtual objects and place them together, including creating new objects by assembling multiple individual objects.&  \cite{elliott} \newline\cite{ochoa} \newline\cite{zhao} 
\\
Basic interaction with objects &Students can select virtual objects and interact with them in different ways. This includes retrieving information about an object in written or spoken form, picking and rotating, zooming in on objects to see more details, and changing an object's color or shape.&  \cite{fittkau} \newline\cite{merino} \newline\cite{vincur}
\\
Instructions &Students have access to a tutorial or to instructions on how to use the VR application and how to perform the learning tasks. The instructions can be given by text, audio, or a virtual agent. This design element does not apply when students have to discover how to use the virtual environment or how to perform learning tasks on their own.&  \cite{zhao} 
\\
Passive observation &Students can look around the virtual environment. This design also applies to applications where users can travel predefined paths and look around while doing so. However, they are not able to move around the environment through them, nor are they able to interact with virtual objects or interact with other users.&  \cite{akbulut} 

\\
  \bottomrule
\end{tabular*}
\end{table*}

We have not found papers that match the other eight categories. It is also worth saying that some papers can fit into over one category. We could identify that the moving around and the user-generated content category are the most popular categories of user-application interaction.

Most applications do not have a direct and deeper interaction, containing decisions and choices that could help the learning process. In addition, only one study \citep{zhao} tried to simulate the requirements process, but it is not specific to the software requirements process.

\subsection{RQ4: What and how are the developer-client-user interactions?}
Software development involves interaction between users and developers. Through this question, we would like to identify how applications have solved this communication resource. 

Despite not approaching the subject directly, \cite{zhao} is the only one that represents the interactions by dealing directly with requirements. We could identify the first station as a customer demanding requirement. We could identify the developer as the student producing the manufactured product. There were no interactions with virtual users.

\subsection{RQ5: Which Unified Model Language (UML) diagrams were adopted and how they were used and designed by students in VR systems?}
With this question, we wanted to identify which and how UML diagrams and models are manipulated in systems with VR.

Only two studies used UML elements and diagrams. By analyzing the attributes in question 2, we could identify that the \cite{tanielu} VR application is a playful version of a class diagram. It allows the user to visualize classes, instances of methods, variables, the entry point of a method, parameters of a method, accessibility of fields, the instance in memory, and its address. \cite{ochoa} presented an interactive software that comprises the connection of floating instances to each other, containing diagrams of use case, state, and data flow.

\cite{ochoa} and \cite{tanielu} gave us different perspectives on how to display UML diagrams to the student. \cite{ochoa} allowed the user to connect UML diagrams interacting with it inside the HMDs, with an approach aiming at a requirement engineering educational tool. On the other hand, \cite{tanielu} exploited VR playfully, but did not allow much interactivity with objects, keeping the application as a fly-through.

\subsection{RQ6: Which were the target hardware of the studies?}
To understand what each application could offer to the student, it is important to understand what each hardware can offer to the developers. In order to answer this research question, \autoref{tab:tab9} shows which HMDs are being used in the selected educational projects.
\begin{table}
  \caption{Hardware used by studies}
  \label{tab:tab9}
   \setlength{\tabcolsep}{10pt} 
\renewcommand{\arraystretch}{1.5}
 \begin{tabular*}{\linewidth}{p{3cm} p{5cm}}
    \toprule
Utilized hardware & Studies\\
    \midrule
   
Google Cardboard &\cite{akbulut} \newline\cite{tanielu}\\
HTC Vive &\cite{merino} \newline\cite{oberhauser} \newline\cite{ochoa} \newline\cite{vincur}  \newline\cite{zhao}\\
Oculus Rift &\cite{elliott} \newline\cite{fittkau} 
\\
  \bottomrule
\end{tabular*}
\end{table}

It is worth mentioning that we contacted by email the authors from \cite{tanielu} and \cite{vincur} studies to clear doubts regarding the hardware used.

\subsection{RQ7: How were the virtual teaching environments in SE or RE evaluated?}

To identify the effectiveness of VR environments for education, it is important to understand how they were evaluated and what their results are. Of the nine selected articles, only seven describe some type of evaluation.

The application of \cite{akbulut} was used by a group of 18 students during 2 hours of practical class per week, while another group of the same size and class schedule was not exposed to the VR class. At the end of the teaching period, a questionnaire with 10 questions was applied to all students in both groups, and when comparing the results, it was noticeable that there was a higher learning gain in the group that had used the technology.

Two students used the application of \cite{oberhauser}, who had to analyze in VR and non-VR a project and its classes, dependencies, sizes and answer a questionnaire about the system usability in a learning context. It was concluded that there was no significant difference between analysis times.

The application of \cite{tanielu}, on the other hand, was evaluated by measuring the degree of confidence of students through questionnaires before and after the use of virtual reality. The study was applied to 17 students in the computer science course and the results were positive.

In \cite{zhao}, twelve engineering students took part in the evaluation. First, they answered a survey with conceptual, analytical, and cognitive questions. Then they tested the game in virtual reality. The same survey was inside the game to verify learning. The paper unpublished the results.

\autoref{tab:tab10} shows the validation strategies used by the selected articles, with Positive (P), Negative (N) or Not Informed (NI) results. following the definitions of \cite{radianti}.

\begin{table}[t]
\begin{FlushLeft}
  \caption{Evaluation strategies}
  \label{tab:tab10}
   \setlength{\tabcolsep}{10pt} 
\renewcommand{\arraystretch}{1.5}
 \begin{tabularx}{\linewidth}{p{2cm} p{3cm} p{0.5cm} p{0.5cm} p{0.5cm}}
    \toprule
Classification & Studies & P & N & NI\\
    \midrule
Empirical, qualitative research
&
\cite{fittkau} \cite{merino} \cite{ochoa}&3&0&-\\

Empirical, quantitative research
 &\cite{akbulut} \cite{oberhauser} \cite{tanielu} \cite{zhao} &2&1&1\\
 
No method explained
&\cite{elliott} \cite{vincur} &0&0&2
\\
  \bottomrule
\end{tabularx}
\end{FlushLeft}
\end{table}

Despite the lack of evaluation of some articles, we can identify that all articles take an empirical approach to evaluate their applications. Most studies base their research on user feelings before and after using a VR application to support teaching a subject. But we could observe that all the selected studies evaluated their applications using a small sample. This makes it difficult to confirm the validity of the experiment.

\section{Discussions and conclusions}
The aim of this paper was to review the literature to identify applications with Virtual Reality in Software and Requirements Engineering education. Virtual reality technology can assist in the development of applications that require user immersion to make learning more effective. One task of a software engineer is to communicate constantly with the stakeholders and teammates. So, one skill that could be improved in the students, with the immersion characteristics of VR, is this communication with these partners. From this conversation, several software models (use cases, classes, etc.) emerged and are improved. These were the features we sought in this systematic review of how they would have been developed. The review conducted of only nine papers found using explicitly VR HMD in software engineering education.

We found it three subjects using simple interactions or immersive visualizations: Programming, Requirement, and Software Architecture. By checking the other project categories of \cite{radianti}, we can note the absence of replication of real worlds and the interaction with other users. We found only one paper \citep{ochoa} that students can create use cases and data flow diagrams. Although \cite{zhao} is not specific to Software Engineering, we selected it because it simulates a very simple software development process, starting with user requirement. Still using the classification of \cite{radianti}, we noted a lack of resources from the learning environment, such as immediate feedback, knowledge testing, virtual reward, the possibility for the teacher to assign and evaluate learning tasks and view student progress.

Almost all papers used expensive technologies as HTC Vive and Oculus Rift. If we want to spread the application or that it is replicable by other researchers, it is more interesting to look for cheaper technologies. Even allowing students to use the learning environments from their homes, or anywhere they went from the educational institution, an alternative is to use mobile devices like Google Cardboard or Apple Glasses.

Finally, many authors consider VR as a valid tool for teaching practically. However, for methodological approaches, we did not find statistical evidence. This happened because the researched works are pilot studies with a small sample and in a specific context. As noted by \cite{radianti}, this may raise doubts about the real effectiveness of validations.

Although virtual reality is being widely used in the teaching of medicine and other engineering courses, remarkably, how this technology still lacks maturity in the teaching of software engineering. Reaffirming what was mentioned earlier by \cite{thiry}, universities are not properly preparing students for software requirements tasks. We believe that simulating real environments with virtual reality can offer the chance for these students to learn as if they were already in their professional lives.

\bibliography{references} 

\end{document}